\documentclass[12pt]{article}
\usepackage{amssymb}

\def\ee{\end{equation}}
\newcommand{\eeq}[1]{\label{#1}\end{equation}}
\def\bea{\begin{eqnarray}}
\def\eea{\end{eqnarray}}

\def \rif  {right--invariant field}

\def\eqn{equation}
\def\tfn{transformation}

\def\cond{condition}

\def\+{{+\!\!\!+}}
\def\-1{^{-1}}
\def\itr{^{-t}}
\def\half{\frac{1}{2}}
\def\unit{{\bf 1}}

\def\real{{\mathbb R}}

\def\dxp{{\partial_+}}
\def\dxm{{\partial_-}}
\def\dpm{{\partial_\pm}}
\def\bndry{|_{\sigma=0,\pi}}

\def\sm{$\sigma$--model}

\def\pltp{Poisson--Lie T--pluralit}
\def\pltd{Poisson--Lie T--dualit}
\def\dd{Drinfel'd double}

\def\wt{\widetilde}
\def\wh{\widehat}
\def\dotplus{\stackrel{{\bf.}}{+}}

\def\cf{{\cal {F}}}

\def\cd{{\mathfrak d}}

\def\cg{{\mathfrak g}}
\def\tcg{{\tilde {\mathfrak g}}}

\def\ttil{\tilde{T}}
\def\htil{\tilde{h}}
\def\that{\widehat{T}}
\def\ghat{\hat{g}}

\def\rpm{\rho_\pm}
\def\rpmg{\rho_\pm(g)}
\def\rpg{\rho_+(g)}
\def\rmg{\rho_-(g)}
\def\rpmtilh{\tilde \rho_\pm(\tilde{h})}
\def\rptilh{\tilde \rho_+(\tilde{h})}
\def\rmtilh{\tilde \rho_-(\tilde{h})}
\def\rphatg{\wh \rho_+(\hat g)}
\def\rmhatg{\wh \rho_-(\hat g)}

\def\gluingmgen{R}
\def\gluingmphi{R_\phi}
\def\gluingmlambda{R_\lambda}
\def\gluingm{R_\rho}
\def\gluingmh{\wh {R_\rho}}
\def\gluingoper{{\cal R}}

\def\PP{p}
\def\QQ{q}
\def\RR{r}
\def\SSS{s}

\begin{document}

\begin{center}

                    \hfill   YITP-07-33 \\

\vskip 1.0in \noindent

{\large \bf{On the Poisson--Lie T--plurality of boundary conditions}}

\vskip .2in

{
Cecilia Albertsson \\
{\em Yukawa Institute for Theoretical Physics,\\
Kyoto University, Kyoto 606-8502, Japan \\
E-mail: cecilia@yukawa.kyoto-u.ac.jp}

\vskip .15in

Ladislav Hlavat\'y and Libor \v Snobl\\
{\em Faculty of Nuclear Sciences and Physical Engineering, \\
Czech Technical University in Prague,\\
B\v rehov\'a 7, 115 19 Prague 1, Czech Republic\\
\vskip .05in E-mail: Ladislav.Hlavaty@fjfi.cvut.cz, Libor.Snobl@fjfi.cvut.cz} }

\end{center}

\vskip .4in

\abstract{Conditions for the gluing matrix defining consistent boundary conditions of two-dimensional
nonlinear $\sigma$--models are analyzed and reformulated. Transformation properties of the right--invariant
fields under Poisson--Lie T--plurality are used to derive a formula for the transformation of the boundary
conditions. Examples of transformation of D--branes in two and three dimensions are presented. We
investigate obstacles arising in this procedure and propose possible solutions.}\vskip6cm

{
Copyright 2008 American Institute of Physics. This article may be downloaded for personal use only. Any other use requires prior permission of the author and the American Institute of Physics.

The following article appeared in J. Math. Phys. 49, 032301 (2008) and may be found at 
\verb#http://link.aip.org/link/?JMP/49/032301#.
}

\newpage

\section{Introduction} \label{Introduction}
T-duality of strings may be realized as a canonical transformation acting on the fields in a two-dimensional nonlinear
$\sigma$--model. This model describes the worldsheet theory of a string propagating on some target manifold equipped
with a background tensor field $\cf_{\mu\nu}$ which is a convenient rearrangement of the metric and the Kalb--Ramond
B--field. For open strings the worldsheet has boundaries, by definition confined to D--branes, hence the action is
subject to boundary conditions. Imposing extra symmetries,  e.g., conformal invariance, restricts these conditions.
They determine the dynamics of the ends of the string, and hence the embedding of D--branes in the target space.
Applying duality transformations yields the dual boundary conditions and hence the geometry of D--branes in the dual
target.

Traditional T-duality requires the presence of an isometry group leaving the $\sigma$--model invariant, a rather
severe restriction. In Poisson--Lie T--duality \cite{klse:dna} isometries are not necessary, provided the two dual
target spaces are both Poisson--Lie group manifolds (or at least Poisson--Lie groups act freely on them) whose Lie
algebras constitute a \emph{Drinfel'd double}. That is, they are maximally isotropic Lie subalgebras in the
decomposition of a Lie bialgebra $\cd=\cg\dotplus\tcg$, where $\tcg \equiv \cg^*$. The background $\cf_{\mu\nu}$ is
related to the Poisson structure on the target manifold and satisfies the \emph{Poisson--Lie condition}, a restriction
that replaces the traditional isometry condition.

Recently the \tfn{} of worldsheet boundary conditions under \pltd y was derived in \cite{are:wsbc}. The key formulae
were \tfn s of left--invariant fields\footnote{The dot denotes matrix multiplication,  $t$ denotes transposition,
$E^{-t}\equiv (E^t)^{-1}$ where $E$ is a general background field in the Lie algebra frame, and $E_0$ is a constant
matrix.}
 \begin{equation} \label{are37}
\tilde L_{\+}^t(\tilde g)=\tilde E\itr(\tilde g) \cdot E_0\itr \cdot E^t(g) \cdot L_{\+}^t(g),
\end{equation}
\begin{equation} \label{are38}\tilde L_{=}^t(\tilde
g)=-(\tilde E(\tilde g))\-1 \cdot E_0\-1  \cdot E(g) \cdot L_=^t(g),
\end{equation} obtained from the canonical transformations derived in \cite{Sfetsos1},\cite{Sfetsos2}. Here $g$ and $\tilde g$ are elements of the groups corresponding to $\cg$ and $\tcg$, respectively, and the subscripts $\+$ and $=$ refer to the worldsheet lightcone coordinates.

Poisson--Lie T--plurality \cite{unge:pltp} is a further generalization of T--duality, where the mutually dual target
spaces do not necessarily belong to the same Lie algebra decomposition of the Drinfel'd double (i.e., they belong to
different \emph{Manin triples}).

In articles \cite{hla:slnbytduality} and \cite{hlahytur} we found classical solutions of \sm s in curved backgrounds
by applying \pltp y \tfn s to flat \sm s. Unfortunately, we were not able to control the boundary conditions necessary
for string solutions in the curved background or, more precisely, to identify the conditions for the flat solution
that transform to suitable conditions in the curved background.

Our goal here is to derive a \tfn{} of boundary \cond s under \pltp y that could enable us to control the boundary
conditions in the transformed \sm. Analogues of the formulae (\ref{are37}) and (\ref{are38}) for \pltp y were derived
in \cite{hlasno:pltpcan} so that we can easily write down the \tfn{} of the boundary \cond s. As the \sm s
investigated in \cite{hla:slnbytduality}, \cite{hlahytur} and other papers of ours are formulated in terms of
right--invariant fields $\dpm g g\-1$ we shall use this formulation here.\footnote{Left--invariant fields were used in
\cite{hlasno:pltpcan}.}

In Section \ref{Elements} we review Poisson--Lie T--plurality and introduce the framework necessary for the subsequent
analysis. In Section \ref{Boundary} we list and discuss the set of boundary conditions required to define consistent
$\sigma$--models, describing them in terms of a gluing matrix. In Section \ref{Transformation} we derive the
T--plurality transformation of the gluing matrix, and show that it does not automatically yield well-defined boundary
conditions on the T--plural side. In Sections \ref{3dimex} and \ref{2dimex} we analyze two explicit examples, one
three-dimensional and one two-dimensional, demonstrating how different D--branes transform under Poisson--Lie
T--plurality. In the process, we discuss the conditions necessary to eliminate any interdependence of the gluing
matrices on coordinates of the different target spaces involved. Finally, Section \ref{Conclusions} contains our
conclusions.

\section{Elements of \pltp y} \label{Elements} The classical action of  the \sm{} under consideration is
\begin{equation} S_{\cf}[\phi]=\int_\Sigma d^2x\,
\partial_- \phi^{\mu}\cf_{\mu\nu}(\phi)
\partial_+ \phi^\nu \label{sigm1} \end{equation}
 where $\cf$ is a tensor on a Lie group $G$ and the functions $ \phi^\mu:\ \Sigma \subset \real^2 \rightarrow \real,\
\mu=1,2,\ldots,{\dim}\,G$ are obtained by the composition $\phi^\mu=y^\mu\circ g $ of a map
 $g:\Sigma\rightarrow G$ and components of a coordinate map $y$ of a neighborhood $U_g$ of an element $g(x_+,x_-)\in G$.
For the purpose of this paper we shall assume that the worldsheet $\Sigma$ has the topology of a strip infinite in
timelike direction, $\Sigma=\langle 0, \pi \rangle \times \real.$

On a Lie group $G$  the tensor $\cf$  can be written
 as \begin{equation} \cf_{\mu\nu}={e_\mu}^a (g)F_{ab}(g)
{e_\nu}^b(g)\label{metorze} \end{equation}
 where ${e_\mu}^a(g)$ are components
of the right-invariant Maurer--Cartan forms ${\rm d}g g^{-1} $ and $F_{ab}(g)$ are matrix elements of bilinear
nondegenerate form $F(g)$ on $\cg$, the Lie algebra of $G$. The action of the \sm{} then reads
\begin{equation} \label{SFg}S_F[g]=\int_\Sigma d^2x\, \rmg \cdot F(g) \cdot \rpg^t \end{equation}
where the right--invariant vector fields $\rpmg$ are given by\footnote{Note that while matrix multiplication is
denoted by dot, for group multiplication we use concatenation.}
\begin{equation}\label{rpm} \rpmg^a\equiv(\dpm g g\-1)^a=\dpm \phi^\mu\, {e_\mu}^a(g),\ \ \ (\dpm g g\-1)=
\rpmg \cdot T, \end{equation} and $T_a$ are basis elements of $\cg$. (Note that $\rpmg$ is written in a condensed
notation, in full detail it would read $\rpm(g(x_+,x_-),\partial_\pm g(x_+,x_-))$ since it is a map $\Sigma\rightarrow
\cg$.)

The \sm s that are transformable under \pltd y can be formulated (see \cite{klse:dna},\cite{kli:pltd}) on a \dd{}
$D\equiv(G|\tilde G)$, a Lie group whose Lie algebra $\cd$ admits a decomposition $\cd=\cg\dotplus\tcg$ into a pair of
subalgebras maximally isotropic with respect to a symmetric ad-invariant nondegenerate bilinear form $\langle\,
.\,,.\,\rangle $. The matrices $F_{ab}(g)$ for the dualizable \sm s are of the form \cite{klse:dna} \begin{equation}
F(g)=(E_0^{-1}+\Pi(g))^{-1}, \ \ \ \Pi(g)=b(g) \cdot a(g)^{-1} = -\Pi(g)^t,\label{Fg}\end{equation}
 where $E_0$ is a constant matrix, $\Pi$ defines the Poisson structure on the group $G$, and $a(g),b(g)$ are submatrices
of the adjoint representation of $G$ on $\cd$. They satisfy
\begin{equation}\label{adgt} g T  g\-1\equiv Ad(g)\triangleright T=a\-1(g) \cdot T,\ \ \ \ g\tilde  T g\-1\equiv
Ad(g)\triangleright \tilde T =b^t(g) \cdot T+ a^t(g) \cdot \tilde T, \end{equation} where $\ttil^a$ are elements of
dual basis in the dual algebra $\tcg$, i.e., $\langle\,T_a ,\,\ttil^b\,\rangle=\delta_a^b $. The matrix $a(g)$ relates
the left-- and right--invariant fields on G, via
\begin{equation}\label{lra} (g\-1 \dpm g)={L_\pm}(g) \cdot T,\ \ \
    {L_\pm}(g)=\rpmg \cdot a(g) .
\end{equation}

The \eqn s of motion of the dualizable \sm s can be written as Bianchi identities for the right--invariant fields
$\rpmtilh$ on the dual algebra $\tcg$ satisfying \cite{kli:pltd}\begin{equation} \label{kl10} \rptilh \cdot
\ttil\equiv(\dxp\htil \htil\-1)=-\rpg \cdot F(g)^t \cdot a\itr(g) \cdot \ttil,
\end{equation}\begin{equation} \label{kl11}
\rmtilh \cdot \ttil\equiv(\dxm\htil \htil\-1)=+\rmg \cdot F(g) \cdot a\itr(g) \cdot \ttil .
\end{equation} This is a consequence of the fact that the \eqn s of motion of the dualizable \sm{} can be
written as the  following \eqn{}s on the \dd{} \cite{klse:dna},
\begin{equation}\label{ddeqm}
    \langle\, \dpm l  l\-1\,,{\cal E^\pm}\,\rangle =0,
\end{equation}
where $l=g \htil$ and ${\cal E^\pm}$ are two orthogonal subspaces in $\cd$. On the other hand, the solution
$g(x_+,x_-)$ of the equations of motion of the action (\ref{SFg}) gives us a flat connection (\ref{kl10}),
(\ref{kl11}), which is therefore locally pure gauge, and the gauge potential $\htil(x_+,x_-)$ is determined up to
right--multiplication by a constant element $\htil_0$. Therefore we find $l(x_+,x_-)=g(x_+,x_-)\cdot \htil(x_+,x_-)$, the
so--called lift of the solution
 $g(x_+,x_-)$ to the \dd{}, determined up to the constant shift
\begin{equation}\label{shift}
l\rightarrow l \htil_0, \qquad \htil_0 \in \tilde G.
\end{equation}

In general, {as was realized already in \cite{klse:dna} and then further developed in \cite{unge:pltp}}, there are
several decompositions (Manin triples) of a \dd. Let $\hat\cg\dotplus\bar\cg$ be another decomposition of the Lie
algebra $\cd$. The pairs of dual bases of $\cg,\tcg$ and $\hat\cg,\bar\cg$ are related by the linear \tfn
\begin{equation}\label{pqrs}
    \left(\matrix{T \cr\tilde T\cr} \right)= \left(\matrix{\PP&\QQ \cr \RR&\SSS \cr} \right) \left(\matrix{\that\cr
\bar T\cr} \right),
\end{equation}
where the duality of both bases requires
\begin{equation}
\left(\matrix{\PP&\QQ \cr \RR&\SSS \cr} \right)\-1=\left(\matrix{\SSS^t&\QQ^t \cr \RR^t&\PP^t \cr}
\right),\end{equation} i.e., \begin{equation} \label{dualbase}\begin{array}{ccr}
\PP \cdot \SSS^t+\QQ \cdot \RR^t&=&\unit,\\
\PP \cdot \QQ^t+\QQ \cdot \PP^t&=&0, \\
\RR \cdot \SSS^t+\SSS \cdot \RR^t&=&0.
\end{array}
\label{KQUSreln2}\end{equation}
 The \sm{} obtained by the plurality \tfn{} is then defined analogously to the original one, namely by substituting
\begin{equation} \wh F(\ghat)=(\wh E_0^{-1}+\wh\Pi(\ghat))^{-1}, \ \ \ \wh\Pi(\ghat)=\wh b(\ghat)
\cdot \wh a(\ghat)^{-1} = -\wh\Pi(\ghat)^t,\label{Fghat}\end{equation}\begin{equation} \label{E0hat} \widehat
E_0=(\PP+E_0 \cdot \RR)\-1 \cdot (\QQ+E_0 \cdot \SSS)=(\SSS^t \cdot E_0-\QQ^t) \cdot (\PP^t-\RR^t \cdot
E_0)\-1\end{equation} into (\ref{metorze}), (\ref{SFg}). Solutions of the two \sm s are related by two possible
decompositions of $l\in D$, namely
\begin{equation}\label{lgh}
    l=g \htil=\ghat \bar h.
\end{equation}
The transformed solution $\ghat$ is determined by the original solution $g(x_+,x_-)$ up to a choice of constant shift
(\ref{shift}).

\section{Boundary conditions and D--branes}
\label{Boundary} The properties of D-branes in the groups $G$ and $\hat G$ can be derived from the so--called gluing
operators $\gluingoper$ and $\wh \gluingoper$, respectively; the number of their $-1$ eigenvalues determines the
number of Dirichlet directions and hence the dimension of the D-branes. Moreover, the explicit form of the operator in
principle yields the embedding of a brane in the target space.

We impose the boundary conditions for open strings in the form of the gluing operator $\gluingoper$ relating the left
and right derivatives of field $g:\Sigma \rightarrow G$ on the boundary of $\Sigma$,
\begin{equation}
\label{bcing} \partial_- g \bndry = \gluingoper \partial_+ g \bndry,\ \ \ \ \ \sigma\equiv x_+ - x_-.
\end{equation}
As we have to work with several choices of coordinates, we denote the matrices corresponding to the operator
$\gluingoper$ in the bases of coordinate derivatives as $\gluingmphi,\gluingmlambda$, etc., e.g.,
\begin{equation}
\partial_-\phi\bndry=\partial_+\phi\cdot \gluingmphi\bndry , \label{cphi}
\end{equation} or
\begin{equation} \partial_-\lambda\bndry=\partial_+\lambda\cdot
\gluingmlambda\bndry , \label{clam}\end{equation} where $\partial_-\phi, \partial_-\lambda$ are row vectors of the
derivatives of the respective coordinates (therefore matrices of operators in our notation may differ
 by a transposition from expressions in other papers). Nevertheless, we suppress the indices $\phi,\lambda$ in expressions valid
in any choice of coordinates, $\gluingmgen$ having the obvious meaning of the matrix of the gluing operator, the tensor $\cf$ is assumed to be expressed in the same coordinates etc.

We define the Dirichlet projector ${\cal Q}$ that projects vectors onto the space normal to the D--brane, which is
identified with the eigenspace of $\gluingoper$ with the eigenvalue $-1$, and the Neumann projector ${\cal N}$, which
projects onto the tangent space of the brane. The corresponding matrices $Q$, $N$ are given by the axioms
\begin{equation}\label{projdef} Q^2=Q,\ \ Q \cdot \gluingmgen = -Q, \ \ N=\unit-Q . \end{equation}

In so--called adapted coordinates $\lambda^\alpha$ (where $\alpha=1,\ldots,\dim G$) the gluing matrix can be written
as \cite{ALZ2}\begin{equation}\label{gluingmlam} \gluingmlambda=\left(\matrix{{\gluingmgen_m}^n&0 \cr 0&-{\delta_i}^j
\cr} \right),\ m,n=1,\ldots,p+1,\ i,j,=p+2,\ldots,\dim G. \end{equation} If the B--field of the model vanishes one can
choose ${\gluingmgen_m}^n={\delta_m}^n$. In such coordinates the terminology becomes clearer as $\lambda^i$ become
coordinates in the Dirichlet directions, $$
\partial_\tau \lambda^i = \frac{1}{2} (\partial_+ +\partial_-) \lambda^i =0 ,$$ whereas $\lambda^m$ are Neumann directions. This
is a traditional misnomer; it is actually a generalization of the Neumann boundary conditions
$$ \partial_\sigma \lambda^m= \frac{1}{2} (\partial_+ -\partial_-) \lambda^m =0$$ to the cases with nonvanishing $B$--field (a better notation might be free boundary conditions, but we shall stick to the traditional ``Neumann'').

To obtain the corresponding boundary conditions written in terms of right--invariant fields $\rpmg$, we must first
express  the gluing operator in the group coordinates $y$  as
$$\gluingmphi=T(y)\cdot \gluingmlambda\cdot T(y)\-1 ,$$
where
$${T(y)_\mu}^\alpha=\frac{\partial\lambda^\alpha}{\partial y^\mu}(y) ,$$
and then transform it into the basis of the Lie algebra of \rif s,
\begin{equation} \gluingm =
e\-1(g)\cdot  \gluingmphi\cdot e(g)=e\-1(g)\cdot T(y)\cdot \gluingmlambda\cdot T(y)\-1\cdot e(g),
\label{hrphi}\end{equation} where $e(g)$ are the right--invariant vielbeins on $G$ introduced in eq.\ (\ref{rpm}). The
boundary conditions may then be expressed in terms of the right--invariant fields, as
\begin{equation}
\label{origbc} \rmg\bndry = \rpg\cdot\gluingm\bndry.
\end{equation}
\medskip

Of course, not  every operator--valued function on the target space, in our case the group $G$, can be interpreted as
a gluing operator, giving consistent boundary conditions for the $\sigma$--model in question. There are several
restrictions on $\gluingoper$, derived, e.g., in \cite{ALZ2}. {We shall briefly recall how these conditions arise and
rewrite them in a slightly more compact but equivalent form.}

First, in order that the adapted coordinates exist in a particular point we must impose
\begin{equation}\label{LieRQ}
\gluingmgen\cdot Q = Q\cdot \gluingmgen\ .
\end{equation}
This is essentially a part of the definition of $Q$; otherwise $Q$ is not fully determined  because to define a
projector we need to specify its image and its kernel. Eq.\ (\ref{projdef}) defines the image of $Q$ to be an
eigenspace of $\gluingmgen$, while eq.\ (\ref{LieRQ}) implies that the kernel is the sum of all the remaining
(generalized) eigenspaces of $\gluingmgen$. On the other hand, condition (\ref{LieRQ}) is a restriction on
$\gluingmgen$ since it tells us that the geometrical\footnote{I.e., the dimension of the eigenspace.} and
algebraic\footnote{I.e., the multiplicity of the root of the characteristic polynomial.} multiplicities of the
eigenvalue $-1$ are equal. If this condition does not hold, one cannot find adapted coordinates (\ref{gluingmlam}) and
the boundary conditions cannot be split into Dirichlet and (generalized) Neumann directions.

The distribution defined by the image of the Neumann projector must be integrable in order to be a tangent space to a
submanifold, i.e., the brane. We find using Frobenius theorem on integrability of distributions that the distribution
must be in involution. When expressed in terms of the matrix $N$ of the Neumann projector this condition reads in any
coordinates,
 \begin{equation}
N_\kappa^{\,\,\,\mu} N_\lambda^{\,\,\,\nu}\partial_{[\mu} N_{\nu]}^{\,\,\,\rho} =0\,. \label{Liepiinteg1}
\end{equation}
In an arbitrary, non--coordinate frame, e.g., when expressed in terms of the right--invariant fields, the condition
(\ref{Liepiinteg1}) appears more complicated. It may in general be expressed using covariant derivatives but for
simplicity we shall use only the coordinate expression (\ref{Liepiinteg1}).

Since our $\sigma$--models are studied with applications to string theory in mind, they are often viewed as gauge
fixed Polyakov actions. This imposes a further constraint on the solutions, in the form of a vanishing stress tensor
$$ {\cal T}_{++} = {\cal T}_{--} = 0$$
(the trace ${\cal T}_{+-}$ vanishes automatically). Enforcing this condition not only in the bulk but also on the
boundary leads to the consistency condition that the gluing operator preserves the metric on the target space, in
other words is orthogonal with respect to the metric. If this condition were not satisfied the $\sigma$--model would
not allow generic string solutions. Explicitly we have
\begin{equation}\label{confcond} \gluingmgen \cdot {\cal G} \cdot \gluingmgen^t = {\cal G} , \end{equation}
where the metric is written as ${\cal G}=(\cf+\cf^t)/2$. Equivalently in the Lie algebra frame $\{ T_a \}$ we express
the metric as $(F+F^t)/2$ and consequently we have
\begin{equation} \gluingm \cdot (F+F^t) \cdot \gluingm^t = (F+F^t). \label{RgRg}\end{equation}

We moreover require that what we identified as Dirichlet and Neumann directions are indeed orthogonal with respect to
the metric on the target space,
\begin{equation}
N \cdot {\cal G} \cdot Q^t =0\, \label{LiepigQ}.\end{equation} When the metric on the target space is positive (or
negative) definite, this is an automatic consequence of (\ref{confcond}). In the pseudo--Riemannian signature it is an
additional constraint weeding out pathological configurations.

Finally, a crucial condition follows from the field variation of the action. Since the boundary conditions should be
such that the variation of the action vanishes not only in the bulk but also on the boundary (that is why we impose
the boundary conditions in the first place) we find by inspection of the boundary term arising in the variation that
under the assumption of locality\footnote{I.e., the integrand itself, not only the integral $\int_{\partial\Sigma}
(\ldots)$, vanishes.} we must impose
$$\delta \phi\cdot  N_\phi \cdot ({\cal F}\cdot  \partial_+ \phi^t - {\cal F}^t \cdot \partial_-\phi^t )\bndry =0, $$
which after the use of eq.\ (\ref{cphi}) becomes
\begin{equation}
\label{bEoMorig} \delta \phi \cdot N_\phi \cdot ({\cal F}  - {\cal F}^t \cdot \gluingmphi^t ) \cdot \partial_+ \phi^t \bndry =0.
\end{equation}
Because $\delta \phi = \delta \phi \cdot N_\phi$ (i.e., $\delta \phi$ is tangent to the brane), and  since $\partial_+
\phi^t$ are not further restricted, we find
\begin{equation}\label{LiepiEpiR}
N \cdot ({\cal F}  - {\cal F}^t \cdot \gluingmgen^t ) =0, \end{equation} which, using eqs.\ (\ref{LieRQ}) and
(\ref{LiepigQ}) as well as the following consequences of the definition of the projectors (\ref{projdef})
\begin{equation}\label{Nprop}
 N \cdot (\unit + \gluingmgen)  =  \unit + \gluingmgen, \qquad  N \cdot (\unit  - \gluingmgen)  = \unit - \gluingmgen -2 Q,
\end{equation}
can be rewritten in an equivalent form originally deduced and used in \cite{ALZ2},
\begin{equation}\label{LiepiEpiRALZ}
N \cdot {\cal F}\cdot  N^t -N\cdot  {\cal F}^t\cdot  N^t \cdot \gluingmgen^t =0.
\end{equation}

 In fact, once we impose condition (\ref{LieRQ}), the pair of conditions (\ref{LiepigQ}) and (\ref{LiepiEpiRALZ})
 are equivalent to the condition (\ref{LiepiEpiR}).
For example, assuming (\ref{LiepiEpiR}) we can establish (\ref{LiepigQ}) as follows,
 $$ 2 N \cdot {\cal G} \cdot Q^t= N \cdot ({\cal F}+{\cal F}^t) \cdot Q^t =  N \cdot ({\cal F} \cdot Q^t -{\cal F}^t \cdot \gluingmgen^t \cdot Q^t) =  N \cdot ({\cal F} -{\cal F}^t \cdot \gluingmgen^t ) \cdot Q^t =0,$$
 where we have used first eq.\ (\ref{projdef}) and then eq.\ (\ref{LiepiEpiR}). Moreover, once we have
established that the condition (\ref{LiepigQ}) holds we know that eqs.\ (\ref{LiepiEpiR}) and (\ref{LiepiEpiRALZ}) are
equivalent.

To summarize, we are lead to the following conditions on a consistent gluing operator $\gluingoper$,
\begin{eqnarray}
\nonumber Q^2=Q, \ N=\unit-Q, \ \gluingmgen \cdot Q = Q \cdot \gluingmgen & = & -Q, \\
\nonumber N_\kappa^{\,\,\,\mu} N_\lambda^{\,\,\,\nu}\partial_{[\mu} N_{\nu]}^{\,\,\,\rho}  & = & 0, \\
\label{allconds} \gluingmgen \cdot {\cal G}\cdot  \gluingmgen^t & = & {\cal G},  \\
\nonumber N \cdot ({\cal F}  - {\cal F}^t \cdot \gluingmgen^t ) & = & 0.
\end{eqnarray}

Next we investigate whether or not these conditions are preserved under \pltp y. As we shall see by investigation of
explicit examples, they are not preserved in general.

\section{\pltp y \tfn s of \rif s and boundary conditions}
\label{Transformation} The derivation of \pltp y \tfn s of left--invariant fields was presented in
\cite{hlasno:pltpcan} but we find it instructive to repeat it  here for the \rif s. In particular we derive the
formulae  generalizing eqs.\ (\ref{are37}) and (\ref{are38}).

Let us write the \rif{} $(\dxp l  l\-1)$ on the \dd{} in terms of $\rpg$ and $\rptilh$,
\begin{eqnarray}\label{rgthil}
    (\dxp l\,  l\-1)=    (\dxp (g \htil) (g \htil)\-1)&=&\rpg\cdot T+\rptilh\cdot g\ttil
    g\-1 \nonumber\\ &=&\rpg\cdot T+\rptilh\cdot \left[b^t(g)\cdot T+a^t(g)\cdot \ttil\right].
\end{eqnarray}
Using the \eqn s of motion (\ref{kl10}) and the formula (\ref{Fg}) for $F(g)$ we get
\begin{eqnarray}\label{rghtil2}
    (\dxp l\, l\-1)  &=&\rpg\cdot T-\rpg\cdot F(g)^t\cdot \left[a\itr(g)\cdot b^t(g)\cdot T+\ttil\right]\nonumber\\
    &=&\rpg\cdot F(g)^t\cdot \left[E_0\itr\cdot T-\ttil\right].
\end{eqnarray}

Similarly, from the decomposition $l=\ghat \bar h$ we get\begin{equation}\label{rghathp}
    (\dxp l\, l\-1)=
   \rphatg\cdot \wh F(\hat g)^t\cdot \left[\wh E_0\itr\cdot \that-\bar T\right].
\end{equation} Substituting the relation (\ref{pqrs}) into eq.\ (\ref{rghtil2}) and comparing
coefficients of $\that$ and $\bar T$ with those in (\ref{rghathp}) we find the \tfn{} of \rif s under \pltp y,
\begin{equation} \label{rphatg} \rphatg=-\rpg \cdot F^{t}(g)
\cdot\left[(E_0^t)\-1\cdot \QQ-\SSS\right]\cdot{\wh F}^{-t}(\hat g). \end{equation} In the same way we can derive
\begin{equation} \label{rmhatg} \rmhatg =\rmg\cdot F(g)\cdot \left[E_0\-1\cdot \QQ+\SSS\right]\cdot \wh F\-1(\hat g). \end{equation}

The formulae (\ref{are37}) and (\ref{are38}) for T--duality are obtained if $\QQ=\unit,\ \SSS=0,\ F(g)=E(g\-1),\
\rpg=-L_=(g\-1),\ \rmg=-L_\+(g\-1)$, in agreement with the alternative version for the \sm{} action used in
\cite{are:wsbc}, \begin{equation}\label{SARE}S_E[g]=\int_\Sigma d^2x\, L_{\+}(g)\cdot E(g)\cdot L_{=}^t(g).
\end{equation}

Substituting  eqs.\ (\ref{rphatg}) and (\ref{rmhatg}) into the gluing condition (\ref{origbc}) we find the T--plural
boundary condition
\begin{equation} \rmhatg\bndry= \rphatg\cdot\gluingmh\bndry , \label{pluralbc}
\end{equation}
where the T--plural gluing matrix is given by \begin{equation} \gluingmh={\wh F}^t(\hat g)\cdot M_-\-1\cdot
F^{-t}(g)\cdot\gluingm(g)\cdot F(g)\cdot M_+\cdot\wh F\-1(\hat g), \label{gluingmh}\end{equation}  and
\begin{equation}\label{Mpm}
 M_+ \equiv \SSS+{E_0}^{-1}\cdot \QQ,\ \ \ M_-\equiv \SSS-{E_0}^{-t}\cdot \QQ.
\end{equation}  Eq.\  (\ref{gluingmh}) defines the transformation of the gluing matrix $\gluingm$  under
Poisson--Lie T--plurality. For Poisson--Lie T--duality, i.e., for $\QQ$=$\RR$=$\unit$, $\PP$=$\SSS$=$0$, the map
(\ref{gluingmh}) reduces (up to transpositions due to the different notation for matrices) to the duality map found in
\cite{are:wsbc},
\begin{equation}\label{Rtransf0} \wt R = -\wt E^{-1} \cdot E_0^{-1}  \cdot E \cdot R
            \cdot E^{-t} \cdot E_0^t \cdot \wt E^t\,.
\end{equation}

An obvious problem is that the transformed gluing matrix $\gluingmh$ may depend not only on $\ghat$ but also on $g$,
i.e., after performing the lift into the double $g \htil=\ghat  \bar h$ it may depend on the new dual group elements
$\bar h\in \bar G$, which contradicts any reasonable geometric interpretation of the dual boundary conditions.
Nevertheless, as we shall see in Section \ref{3dimex}, if $g$ and $\ghat$ represent the maps $\Sigma \rightarrow G$
and $\Sigma \rightarrow \hat G$ related by the plurality \tfn{}, the boundary conditions (\ref{origbc}) and
(\ref{pluralbc}) are equivalent in the sense that they result in the same conditions on arbitrary functions (see e.g.\
(\ref{d151}) ) occurring in solutions of the Euler--Lagrange equation of the action (\ref{SFg}).

The T--plural counterparts of the Dirichlet and Neumann projectors may be consistently introduced in the same manner
as for the T-dual case \cite{are:wsbc}, letting the relations $\wh \gluingmgen\cdot \wh Q =\wh Q \cdot \wh \gluingmgen
= -\wh Q$ and $\wh N=\unit-\wh Q$  define $\wh Q$ and $\wh N$ on $\wh G$. When the conditions (\ref{allconds}) are
satisfied also for $\wh \gluingmgen,\, \wh Q,\,\wh N$, then given a nonlinear $\sigma$--model on $G$ with well-defined
boundary conditions, we find a $\sigma$--model on $\wh G$ with well-defined boundary conditions.

The conformal condition (\ref{confcond}) is preserved under the \pltp y, i.e., eq.\ (\ref{RgRg}) implies
\begin{equation} \wh{\gluingmphi} \cdot \wh {\cal G} \cdot\wh{\gluingmphi}^t = \wh{\cal G},\
\ \ \ \gluingmh  \cdot\wh G(g) \cdot \gluingmh^t = \wh G(g). \label{pRgRg}\end{equation} This is seen by using eqs.\
(\ref{RgRg}) and (\ref{gluingmh}), as well as the identities
\begin{equation} F(g)^{-t}\cdot G(g)\cdot
F(g)^{-1}=E_0^{-1}+ E_0^{-t}=M_\pm\cdot(\wh E_0^{-1}+ \wh E_0^{-t})\cdot M_\pm^{t},\end{equation} which follow from
eqs.\ (\ref{dualbase})--(\ref{E0hat}).

Imposing the condition (\ref{LiepiEpiR}) on the T--plural model, and working in the basis of right--invariant fields,
we may substitute eq.\ (\ref{gluingmh}) in the left--hand side of eq.\ (\ref{LiepiEpiR}), to obtain
\begin{equation}\label{lhsLiepiEpiR}
\wh  N \cdot (\wh F - \wh F^t \cdot\gluingmh^t) = \wh  N \cdot ( \wh F - \wh F^t \cdot\wh F^{-t}\cdot
(\SSS+E_0^{-1}\cdot \QQ)^t \cdot C^t \cdot (\SSS-E_0^{-t} \cdot \QQ)^{-t} \wh F ),
\end{equation}
where we have defined $C\equiv F^{-t}(g)\cdot\gluingm(g)\cdot F(g)$. This simplifies to
$$ \wh  N \cdot \left( (\SSS-E_0^{-t} \cdot \QQ)^{t} - (\SSS+E_0^{-1} \cdot \QQ)^t \cdot C^t  \right) \cdot (\SSS-E_0^{-t}\cdot \QQ)^{-t} \cdot \wh F. $$
The last two terms are by construction regular matrices and can be omitted while investigating when the expression
(\ref{lhsLiepiEpiR}) vanishes. Consequently, the T--plural version of condition (\ref{LiepiEpiR}) has the form
\begin{equation}\label{pluralLiepiEpiR}
\wh  N \cdot \left( (\SSS-E_0^{-t} \cdot  \QQ)^{t} - (\SSS+E_0^{-1}\cdot \QQ)^t \cdot C^t  \right)=0.
\end{equation}

To gain a better understanding of eq.\ (\ref{pluralLiepiEpiR}), consider the particular case of originally
purely Neumann boundary conditions, i.e., free endpoints. In this case $\gluingm(g) = F^{t}(g)\cdot
F^{-1}(g)$, i.e., $C=\unit$, and the transformation (\ref{gluingmh}) is well--defined { (i.e., $\gluingmh$
is function of $\hat g$ only)} on any of the groups in any decomposition of the \dd{}. This means that any
T--plural $\wh \gluingmgen$ depends on the coordinates on the respective group $\hat G$ only. In this case
the condition (\ref{pluralLiepiEpiR}) further simplifies to
\begin{equation}\label{dualLiepiEpiR}
\wh  N \cdot \QQ^t=0,
\end{equation}
where again regular matrices have been omitted in the product.  We conclude that in the case of Poisson--Lie
T--duality, where $\QQ=\unit$,
 the dual gluing operator  satisfies condition (\ref{LiepiEpiR}) only if
 it is completely Dirichlet, in which case the dual version of (\ref{LiepiEpiR}) is trivially satisfied.

A possible solution to this problem, considered already in \cite{klse:pltdosdb}, comes from the fact that
the condition (\ref{LiepiEpiR}) is modified if the endpoints of the string are electrically charged.  Let us
modify the action by boundary terms
\begin{equation}
S_{\cf}[\phi] \rightarrow S_{\cf}[\phi] + S_{boundary}[\phi]
\end{equation}
where
\begin{equation}
S_{boundary}[\phi] = q_0 \int_{\sigma=0} A_\mu \frac{\partial \phi^\mu}{\partial \tau} {\rm d}\tau +q_\pi \int_{\sigma=\pi} A_\mu \frac{\partial \phi^\mu}{\partial \tau} {\rm d}\tau
\end{equation}
corresponds to electrical charges $q_0,q_\pi$ associated with the two endpoints of the string interacting with electric field(s) present on the respective D--branes. In order to make the following derivation easily comprehensible let us assume that the potential $A_\mu$ can be in an arbitrary but smooth way extended to the neighborhood of the respective brane\footnote{generalization to the case when this is not possible will be explained below} and denote the field strength of the potential $A_\mu$ by\footnote{recall that $y^\mu$ are coordinates on $G$ and $\phi^\mu=y^\mu\circ g$}
\begin{equation}\label{defdelta}
 \Delta_{\mu\nu}= \frac{1}{2} \left( \frac{\partial A_\nu}{\partial y^\mu} - \frac{\partial A_\mu}{\partial y^\nu} \right), \qquad {\rm i.e.} \quad \Delta={\rm d} A.
\end{equation}
Consequently, the equations of motion in the bulk obtained by the variation of the action are left unchanged but
we find on the boundary
\begin{equation}
\delta \phi \cdot N_\phi \cdot ({\cal F}  - {\cal F}^t \cdot \gluingmphi^t + q_0 \Delta \cdot (1+\gluingmphi^t) ) \cdot \partial_+ \phi^t |_{\sigma=0} =0
\end{equation}
together with
\begin{equation}
\delta \phi \cdot N_\phi \cdot ({\cal F}  - {\cal F}^t \cdot \gluingmphi^t - q_\pi \Delta \cdot (1+\gluingmphi^t) ) \cdot \partial_+ \phi^t |_{\sigma=\pi} =0
\end{equation}
instead of (\ref{bEoMorig}).
Therefore, by similar arguments as before we find the following conditions instead of (\ref{LiepiEpiR})
\begin{eqnarray}
\nonumber N \cdot ({\cal F}  - {\cal F}^t \cdot \gluingmgen^t + q_0 \Delta \cdot (1+\gluingmgen^t) )|_{\sigma=0} & = & 0, \\
\nonumber  N \cdot ({\cal F}  - {\cal F}^t \cdot \gluingmgen^t - q_\pi \Delta \cdot (1+\gluingmgen^t) )|_{\sigma=\pi} & = & 0 .
\end{eqnarray}
Because these conditions should hold irrespective of which of the two endpoints lies on the considered
brane (i.e. on any given brane a string may begin, end or both)  we see that the endpoints are oppositely
charged (and by proper choice of convention for $A_\mu$ we set the charge to unity)
\begin{equation}
 q_0=-q_\pi=1.
\end{equation}
That means that the condition (\ref{LiepiEpiR}) modified  by the presence of electric charge at the endpoints reads
\begin{equation}\label{LiepiEpiRA}
 N \cdot \left( ({\cal F}+\Delta ) - ({\cal F}+\Delta )^t \cdot \gluingmgen^t \right)  =  0.
\end{equation}
In fact, recalling eq. (\ref{Nprop}) and writing
\begin{equation}\label{NDeltaN}
  N \cdot  \left(\Delta-\Delta^t \cdot \gluingmgen^t \right) = N \cdot  \Delta \cdot \left(\unit+\gluingmgen^t \right) = N \cdot  \Delta \cdot N^t \cdot \left(\unit+\gluingmgen^t \right)
\end{equation}
 we see that only derivatives of $A_\mu$ along the brane are relevant in the variation of the action $S_{\cf}[\phi] + S_{boundary}[\phi]$, i.e. the resulting condition (\ref{LiepiEpiRA}) doesn't depend on the way we extend $A_\mu$ outside the brane.  If such an extension is impossible the definition (\ref{defdelta}) of $\Delta$ is obviously meaningless and must be corrected in the following way. We introduce the embedding $\iota$ of the brane ${\cal B}$
$$ \iota: {\cal B} \rightarrow G, \qquad {\cal B} \simeq \iota({\cal B}) \subset G$$
and construct the electric field on the brane as
\begin{equation}
 \Delta_{\cal B} = {\rm d}_{\cal B} A \in \Omega^2({\cal B}).
\end{equation}
Then we may pointwise extend $\Delta_{\cal B}|_p$ to a two--form $\Delta|_{\iota(p)}$ with values in $\Omega^2_{\iota(p)}(G)$ (i.e. a two--form on $G$ in the point ${\iota(p)}$)
\begin{equation}\label{defdeltaext}
 \Delta(V,W)|_{\iota(p)} = \Delta_{\cal B} \left( {\cal N} (V), {\cal N} (W)  \right)|_{p}, \qquad p \in {\cal B},\; V,W \in T_{\iota(p)} G
\end{equation}
(where the natural identification $T_p {\cal B}\simeq\iota_*(T_p {\cal B})={\rm Im}({\cal N})|_{\iota(p)}$ is assumed).
With this understanding in mind the condition (\ref{LiepiEpiRA}) remains the same as before but supplemented by a consequence of (\ref{defdeltaext})
\begin{equation}\label{deltarestricts}
\Delta=N.\Delta.N^t.
\end{equation}

Consequently, even if the target group $G$ is foliated by D--branes, and $\Delta$ constructed as a collection of $\Delta$'s on different branes may be well--defined and smooth on $G$ (or its open subset), $\Delta$ may nonetheless
not be closed  -- only its restrictions $\Delta|_{\cal B}$ to the respective branes need to be closed in order to allow the potential $A_\mu$ along the brane.

In the following we shall use the condition (\ref{LiepiEpiRA}) to look for suitable background electric field strength $\Delta$ such that the boundary equations of motion are satisfied in the transformed models. Taking into account (\ref{NDeltaN}) we see that (\ref{LiepiEpiRA}) determines $\Delta=N.\Delta.N^t$ uniquely and generically smoothly (except when $N$ changes rank). The self--consistency of such a procedure of course requires that $\Delta$ found in this way is closed along the branes, i.e.
\begin{equation}\label{cor_closure}
 {N_\kappa}^\nu {N_\lambda}^\rho {N_\mu}^\sigma \partial_{[\nu} \Delta_{\rho \sigma]} =0
\end{equation}
 and hence\footnote{ up to possible topological obstructions which we shall neglect here} gives rise to the potential $A_\mu$.

We should note that the case of free endpoints, i.e., purely Neumann boundary condition $\gluingm(g) = F^{t}(g)\cdot F^{-1}(g)$, was investigated in \cite{klse:pltdosdb}. The approach used there was based on symplectic geometry and it was shown that the Poisson--Lie
T--dual configuration corresponds to D--branes as symplectic leaves of the Poisson structure on the dual group $\tilde
G$ (once one fixes one end of the dual string at the origin of $\tilde G$ using the freedom of a constant shift
(\ref{shift})) and that the correction $\Delta$ in this case exists and is obtained from Semenov--Tian--Shansky symplectic form on the \dd{} as symplectic form on the symplectic leaves and is therefore closed along the branes. Those results are in accord with the analysis here. Also it is clear from the conclusions of \cite{klse:pltdosdb} that in this particular case the integrability condition (\ref{Liepiinteg1}) is automatically satisfied on the dual since the symplectic leaves are submanifolds.

\section{Three--dimensional example}\label{3dimex}
As mentioned in  Section \ref{Introduction}, there are several explicitly solvable \sm s whose solutions are related
by \pltp y. We can construct their gluing matrices corresponding to D-branes and check the equivalence of eqs.\
(\ref{origbc}) and (\ref{pluralbc}). Here we present a three--dimensional example, where one of the solutions is flat
with vanishing B--field while the T--plural one is curved and torsionless. They are given by a six--dimensional \dd{}
with decompositions into, on the one hand, the $Bianchi\,5$ and $Bianchi\,1$ algebras, and on the other hand,  the
$Bianchi\,6_0$ and $Bianchi\,1$ algebras. On $Bianchi\,5$ the background is given by
\begin{equation}\label{E0}
   E_0 =F(g)= \left(\matrix{ 0  & 0&  {\kappa }
   \cr 0& {\kappa }  & 0
   \cr \kappa   &0  & 0 \cr } \right),\ \  \ \kappa \in \real.
\end{equation} The right--invariant vielbein in a convenient parameterization $g=g(y^\mu)$
of the solvable group corresponding to $Bianchi\,5$  is
\begin{equation}\label{eR}
   e(g)= \left(\matrix{ 1  & 0&  0
   \cr 0& e^{- y^{1}}  & 0
   \cr 0 &0  & e^{- y^{1}} \cr } \right) ,
\end{equation} so that the tensor field of the conformal \sm{} that lives on
this group reads
\begin{equation}\label{F51}
{\cal F}_{\mu\nu}(y)=\left(
                \begin{array}{ccc}
                  0 & 0 & \kappa e^{- y^{1}} \\
                  0 & \kappa e^{-2 y^{1}} & 0 \\
                  \kappa e^{- y^{1}} & 0 & 0
                \end{array}
              \right).
\end{equation}
The metric of this model is indefinite and flat. The general solution of the \eqn s of motion is
\cite{hlahytur}\begin{eqnarray}\label{solphi1}
\begin{array}{cl}
{\phi}^1(x_+,x_-)=&-\ln(-W_1-Y_1),\\ \\
{\phi}^2(x_+,x_-)=&-\frac{W_2+Y_2}{W_1+Y_1},\\ \\
{\phi}^3(x_+,x_-)=&W_3+Y_3+\frac{(W_2+Y_2)^2}{2(W_1+Y_1)},
\end{array}
\end{eqnarray}where $W_j=W_j(x_+),\ Y_j=Y_j(x_-)$ are arbitrary
functions.

The \sm{} related to that on $Bianchi\,5$ by \pltp y lives on the solvable group corresponding to $Bianchi\,6_0$, and
its tensor field obtained from
\begin{equation}\label{fhatg601}  \wh{E_0}= \wh F(\hat g)=\left(
\begin{array}{ccc}
 \frac{1}{\kappa } & \frac{1}{\kappa } & \frac{\kappa }{2} \\
 \frac{1}{\kappa } & \frac{1}{\kappa } & -\frac{\kappa }{2} \\
 \frac{\kappa }{2} & -\frac{\kappa }{2} & 0
\end{array}
\right)
\end{equation}
and
\begin{equation}\label{eRhat}
   \wh e(\hat g)= \left(\matrix{ \cosh {\widehat y^{3}}  & -\sinh {\widehat y^{3}}&  0
   \cr -\sinh {\widehat y^{3}}& \cosh {\widehat y^{3}}  & 0
   \cr 0 &0  & 1 \cr } \right)
\end{equation} reads
\begin{equation}\label{F601}
\widehat {{\cal F}_{\mu\nu}}(\widehat y)=\left(
 \begin{array}{ccc}
    \frac{1}{{\kappa}}e^{-2\widehat y^{3}} & \frac{1}{{\kappa}}e^{-2\widehat y^{3}} & \frac{{\kappa}}{2}e^{\widehat y^{3}}
    \\ \\
    \frac{1}{{\kappa}}e^{-2\widehat y^{3}} & \frac{1}{{\kappa}}e^{-2\widehat y^{3}} & -\frac{{\kappa}}{2}e^{\widehat y^{3}}
    \\ \\
    \frac{{\kappa}}{2}e^{\widehat y^{3}} & -\frac{{\kappa}}{2}e^{\widehat y^{3}} & 0 \\
  \end{array}
\right).
\end{equation}
The Ricci tensor of this metric is nontrivial so that the background is curved but has zero Gauss curvature.

The transformation (\ref{pqrs}) between the bases of decompositions of the Lie algebra of the \dd{} into
$Bianchi\,5\dotplus Bianchi\,1$ and $Bianchi\,6_0\dotplus Bianchi\,1$ is given by the matrix
\begin{equation} \left(\matrix{\PP&\QQ \cr \RR&\SSS \cr} \right) =\left(
\begin{array}{cccccc}
 0 & 0 & -1 & 0 & 0 & 0 \\
 0 & 0 & 0 & 1 & 1 & 0 \\
 -1 & 1 & 0 & 0 & 0 & 0 \\
 0 & 0 & 0 & 0 & 0 & -1 \\
 \frac{1}{2} & \frac{1}{2} & 0 & 0 & 0 & 0 \\
 0 & 0 & 0 & -\frac{1}{2} & \frac{1}{2} & 0
\end{array}
\right), \label{C51to601}\end{equation} and the coordinate \tfn{} on the \dd{} that follows from this reads (see
\cite{hlahytur})

\begin{eqnarray}\label{hatphi1}
\begin{array}{rl}
\widehat{y}^1=&-y^3+\frac{1}{2}\tilde{h}_2,  \\  \\
\widehat{y}^2=&y^3+\frac{1}{2}\tilde{h}_2,  \\  \\
\widehat{y}^3=&-y^1,   \end{array}
\end{eqnarray}
\begin{eqnarray}\label{hprime}
\begin{array}{rl}
\bar h_1=&-\frac{1}{2}\tilde{h}_3+y^2,  \\ \\
\bar h_2=&\frac{1}{2}\tilde{h}_3+y^2, \\   \\
\bar h_3=&-\tilde{h}_1+\tilde{h}_2\,y^2 ,
\end{array}
\end{eqnarray}
where $y,\,\tilde h,\,\widehat y,\,\bar h$ are coordinates on the respective subgroups $G,\,\tilde G,\,\widehat
G,\,\bar G$ that correspond to the different decompositions of the \dd. Inserting  eqs.\ (\ref{solphi1}) and the
solution of eqs.\ (\ref{kl10}), (\ref{kl11}) into eqs.\ (\ref{hatphi1}), we obtain the solution \cite{hlahytur} of the
equations of motion for the \sm{} in the curved background given by $\widehat F$,
\begin{eqnarray}\label{solhatphi1}
\begin{array}{cl}
\widehat{\phi}^1(x_+,x_-)=&\frac{1}{2}\kappa\left[{Y_1}({x_-}){W_2}({x_+})-{Y_2}({x_-}){W_1}({x_+})\right]
-\left[{W_3}({x_+})+{Y_3}({x_-})\right]\\
\\&-\frac{1}{2}{\frac{\left[{W_2}({x_+})+{Y_2}({x_+} )\right
]^{2}}{\left({W_1}({x_+})+{Y_1}({x_-})\right]}}+\frac{1}{2}\kappa(\alpha({x_+})+\beta({x_-})),\\ \\
\widehat{\phi}^2(x_+,x_-)=&\frac{1}{2}\kappa\left[{Y_1}({x_-}){W_2}({x_+})-{Y_2}({x_-}){W_1}({x_+})\right]+\left[{W_3}({x_+})+{Y_3}({x_-})\right]\\
\\&+\frac{1}{2}{\frac {\left ({W_2}({x_+})+{Y_2}({x_-} )\right
)^{2}}{{W_1}({x_+})+{Y_1}({x_-})}}+\frac{1}{2}\kappa (\alpha({x_+})+\beta({x_-})),
\\ \\
\widehat{\phi}^3(x_+,x_-)=&\ln (-{W_1}({x_+})-{Y_1}({x_-})),
\end{array}
\end{eqnarray}where  $\alpha,\ \beta$ satisfy (primes denote differentiation)
\begin{eqnarray}\label{beta}
\alpha\,'&=&W_1W_2'-W_2W_1',\\
\beta\,'&=&Y_2Y_1'-Y_1Y_2' . \nonumber
\end{eqnarray}

\subsection{D--branes}\label{Dbranes} In the
following we analyze examples of D--branes for which the adapted coordinates $\lambda^\alpha$ of the flat model are
equal to those that bring the metric of the flat model to the diagonal form
$$F_{kl}(\lambda)=\left(
                \begin{array}{ccc}
                  -\kappa & 0 & 0 \\
                  0 & \kappa & 0 \\
                  0& 0 & \kappa
                \end{array}
              \right),$$namely\begin{eqnarray}
 \nonumber \lambda^1(y)&=& \lambda^1_0-\frac{1}{\sqrt{2}}\left[y^{3}+\frac{1}{2}(y^{2})^{2}e^{-y^1}+e^{-y^1}\right], \\
  \lambda^2(y)&=&\lambda^2_0+  y^{2}e^{-y^1}, \\
\nonumber \lambda^3(y)&=&\lambda^3_0+ \frac{1}{\sqrt{2}}\left[y^{3}+\frac{1}{2}(y^{2})^{2}e^{-y^1}-e^{-y^1}\right].
\end{eqnarray} In these coordinates the gluing matrices $\gluingmlambda$ by assumption
become diagonal \cite{ALZ2}.
\begin{itemize}
    \item D2-branes. The Dirichlet projector is zero (and the Neumann projector is the identity)
in this case and as the tensor $\cf$ is symmetric it follows from eq.\ (\ref{LiepiEpiR}) that the gluing matrices are
    \begin{equation} \gluingmlambda=\gluingmphi=\gluingm=\left(
                \begin{array}{ccc}
                  1 & 0 & 0 \\
                  0 & 1 & 0 \\
                  0& 0 & 1 \\
                \end{array}
              \right).
\end{equation}  The conditions (\ref{allconds}) are trivially satisfied. The condition (\ref{origbc}), or equivalently (\ref{clam}), then gives the
boundary conditions for the solution (\ref{solphi1}),
\begin{equation} W_j'(x_+)\bndry=Y_j'(x_-)\bndry,\ \ j=1,2,3.\label{d251}\end{equation}

From eq.\ (\ref{gluingmh}) we get \begin{equation} \gluingmh=\wh{\gluingmphi}=\left(
                \begin{array}{ccc}
                  0 & -1 & 0 \\
                  -1 & 0& 0 \\
                  0& 0 & 1 \\
                \end{array}
              \right).
\label{D2Rhat}\end{equation} This matrix has eigenvalues (-1,1,1) and the eigenvector corresponding to the eigenvalue
   -1 is spacelike in the (curved) metric
(\ref{F601}) so that the D2--brane is transformed to a D1--brane. The Dirichlet projector obtained from eqs.\
(\ref{projdef}) and (\ref{LieRQ}) is\begin{equation} \wh Q=\left(
                \begin{array}{ccc}
                  \frac{1}{2} & \frac{1}{2}  & 0 \\
                  \frac{1}{2} & \frac{1}{2} & 0 \\
                  0& 0 & 0 \\
                \end{array}
              \right),
\end{equation}and the conditions (\ref{allconds}) are satisfied for the  matrix  (\ref{D2Rhat}). Using
 eqs.\ (\ref{solhatphi1})
and (\ref{d251}) one can verify that
\begin{equation}
\partial_-\wh\phi\bndry=\partial_+\wh\phi\cdot \wh{\gluingmphi}\bndry, \end{equation}
which is equivalent to eq.\ (\ref{pluralbc}). Note that unlike the D1-branes and D0-branes discussed below, in this
case neither the matrix $\gluingm$ nor $\gluingmh$ depends on elements of the groups $G$ and $\hat G$.

\item D1-branes. We have chosen the branes as coordinate planes of the flat coordinates, i.e.,
    \begin{equation} \gluingmlambda=\left(
                \begin{array}{ccc}
                  1 & 0 & 0 \\
                  0 & 1 & 0 \\
                  0& 0 & -1 \\
                \end{array}
              \right),
\end{equation}which in $y$--coordinates gives the $y$--dependent gluing matrix
\begin{equation} \gluingmphi=\left(
\begin{array}{ccc}
 \frac{(y^2)^2}{2} & \frac{1}{2} y^2
   \left[(y^2)^2-2\right] & -\frac{1}{4} e^{-y^1}
   \left[(y^2)^2-2\right]^2 \\
 -y^2 & 1-(y^2)^2 & \frac{1}{2} e^{-y^1} y^2
   \left[(y^2)^2-2\right] \\
 -e^{y^1} & -e^{y^1} y^2 & \frac{(y^2)^2}{2}
\end{array}
\right). \end{equation}The Dirichlet projector obtained from
 eqs.\ (\ref{projdef}) and (\ref{LieRQ}) is\begin{equation}
Q= \left(
\begin{array}{ccc}
 \frac{1}{4} \left[2-(y^2)^2\right] & \frac{1}{4} \left[2 (y^2)-(y^2)^3\right] & \frac{1}{8} e^{-y^1} \left[(y^2)^2-2\right]^2
   \\
 \frac{1}{2}(y^2)^2 & \frac{1}{2}(y^2)^2 & -\frac{1}{4} e^{-y^1}
   (y^2) \left[(y^2)^2-2\right] \\
 \frac{1}{2}{e^{y^1}} & \frac{1}{2}
   e^{y^1} (y^2) & \frac{1}{4} \left[2-(y^2)^2\right]
\end{array}
\right),
\end{equation} and the conditions (\ref{allconds}) are satisfied. The
condition (\ref{origbc}) then gives
\begin{eqnarray}
  W_1'(x_+)\bndry&=&Y_3'(x_-)\bndry\nonumber, \\
  W_2'(x_+)\bndry&=&Y_2'(x_-)\bndry  \label{d151}, \\
  W_3'(x_+)\bndry&=&Y_1'(x_-)\bndry\nonumber.
\end{eqnarray}

From eq.\ (\ref{gluingmh}) we obtain $\gluingmh$ and $\wh{\gluingmphi}$, which however are too complicated
to be displayed here. The matrix $\wh{\gluingmphi}$ depends on the coordinates on both $\wh G$ { and $G$ and
consequently on $\bar G$}, nevertheless, we have checked that the boundary condition (\ref{pluralbc}) for
the solution (\ref{solhatphi1}) implies again the relations (\ref{d151}). In this sense the conditions
(\ref{origbc}) and (\ref{pluralbc}) are equivalent.

The eigenvalues of $\wh \gluingmphi$ are $1,-1+(y^2)^2\pm\sqrt{(y^2)^4-2 (y^2)^2}$  so that for $y^2\neq 0$ the
projectors are $\wh Q=0,\ \wh N=\unit$, and the condition (\ref{LiepiEpiR}) is not satisfied.

On the other hand, the hypersurface $y^2=0$ does not coincide with a D1--brane in the original model since
the tangent vector $\partial_{y^2} |_{y^2=0}$ is Neumann. Consequently, if at a given time the endpoint of a
string is located at $y^2=0$, it might not stay there at later times. We conclude that in this case the
transformed D--brane configuration is not well--defined { due to the dependence of $\gluingmh$ on the
coordinates on $\bar G$}.
\item D0-branes. We choose
    \begin{equation} \gluingmlambda=\left(
                \begin{array}{ccc}
                  1 & 0 & 0 \\
                  0 & -1 & 0 \\
                  0& 0 & -1 \\
                \end{array}
              \right),
\end{equation}so that\begin{equation} \gluingmphi=\left(\begin{array}{ccc}
 \frac{(y^2)^2}{2} & \frac{(y^2)^3}{2}+y^2 &
   -\frac{1}{4} e^{-y^1} \left[(y^2)^2+2\right]^2 \\
 -y^2 & -(y^2)^2-1 & \frac{1}{2} e^{-y^1} y^2
   \left[(y^2)^2+2\right] \\
 -e^{y^1} & -e^{y^1} y^2 & \frac{(y^2)^2}{2}
\end{array}
\right). \end{equation}The Dirichlet projector is
\begin{equation} Q= \left(
\begin{array}{ccc}
 \frac{1}{4} \left[2-(y^2)^2\right] & \frac{1}{4} \left[-(y^2)^3-2 (y^2)\right] &  \frac{1}{8} e^{-y^1} \left[(y^2)^2+2\right]^2
   \\
 \frac{(y^2)}{2} & \frac{1}{2}
   \left[(y^2)^2+2\right] & -\frac{1}{4} e^{-y^1}
   (y^2) \left[(y^2)^2+2\right] \\
 \frac{1}{2}{e^{y^1}} & \frac{1}{2} e^{y^1} (y^2) & \frac{1}{4} \left[2-(y^2)^2\right]
\end{array}
\right)
\end{equation}
and the conditions (\ref{allconds}) are satisfied. The condition (\ref{origbc}) yields \begin{eqnarray}
  W_1'(x_+)\bndry&=&-Y_3'(x_-)\bndry\nonumber, \\
  W_2'(x_+)\bndry&=&-Y_2'(x_-)\bndry \label{d051},\\
  W_3'(x_+)\bndry&=&-Y_1'(x_-)\bndry\nonumber.
\end{eqnarray}

The matrix $\wh \gluingmphi$  is again rather complicated and depends on the coordinates of both $G$ and $\wh G$, but
once again using eqs.\ (\ref{solhatphi1}) and (\ref{d051}) one can verify that the conditions (\ref{origbc}) and
(\ref{pluralbc}) are equivalent in the sense explained above.

The eigenvalues of $\wh \gluingmphi$ are $-1,1+(y^2)^2\pm\sqrt{(y^2)^4+2 (y^2)^2}$ and the  Dirichlet projector $\wh
Q$ obtained from eqs.\  (\ref{projdef}) and (\ref{LieRQ}) reads \begin{equation}\wh Q=\left(
\begin{array}{ccc}
 \frac{1}{4} & -\frac{1}{4} &  \frac{e^{2 y^1+\wh{y^3}}}{2 (y^2)^2+4} \\
 -\frac{1}{4} & \frac{1}{4} & -\frac{e^{2 y^1+\wh{y^3}}}{2 (y^2)^2+4} \\
\frac{1}{4} e^{-2 y^1-\wh{y^3}} \left((y^2)^2+2\right) & -\frac{1}{4} e^{-2 y^1-\wh{y^3}} \left((y^2)^2+2\right) &
   \frac{1}{2}
\end{array}
\right) . \label{whqd0} \end{equation} { Due to (\ref{hatphi1}-\ref{hprime}), namely $y^2=\frac{1}{2} (\bar
h_1+\bar h_2) $, the projector $\wh Q$ depends both on $\wh G$ and $\bar G$.  } The conditions
(\ref{allconds}) are again satisfied only for $y^2=0$ but now the tangent vector $\partial_{y^2} |_{y^2=0}$
is Dirichlet. We can therefore consistently restrict ourselves in the original model to D0--branes inside
the hypersurface $y^2=0$. Their plural counterparts are given by a gluing matrix of the form
\begin{equation} \label{hathphid0}\wh {\gluingmphi}=\left( \matrix{ \frac{1}{2}  &  \frac{1}{2}
      & -\frac{1}{2}{e^{- {\wh y}^3}}
      \cr \frac{1}{2}    &  \frac{1}{2}    & \frac{1}{2}{e^
     {- {\wh y}^3}} \cr -e^
    { {\wh y}^3} & e^
   { {\wh y}^3} & 0 \cr  }\right) , \end{equation}
where we have used the coordinate \tfn{} (\ref{hatphi1}). Its eigenvalues are $(-1,1,1)$ and the eigenvector
corresponding to the eigenvalue $-1$ is spacelike so that the matrix (\ref{hathphid0}) defines a D1--brane in the dual
model.

\item D(-1)-branes. We have
\begin{equation} \gluingmlambda=\gluingmphi=\gluingm=\left(
                \begin{array}{ccc}
                  -1 & 0 & 0 \\
                  0 & -1 & 0 \\
                  0& 0 & -1 \\
                \end{array}
              \right).
\end{equation} The Dirichlet projector is the identity in this case so that the conditions (\ref{allconds}) are trivially satisfied. The
 condition (\ref{origbc})
then gives the boundary conditions for the solution (\ref{solphi1}),
\begin{equation}
W_j'(x_+)\bndry=-Y_j'(x_-)\bndry,\ \ j=1,2,3.\label{dm151}
\end{equation}

From eq.\ (\ref{gluingmh}) we find \begin{equation} \label{hathphidm1}\gluingmh=\wh{\gluingmphi}=\left(
                \begin{array}{ccc}
                  0 & 1 & 0 \\
                  1 & 0& 0 \\
                  0& 0 & -1 \\
                \end{array}
              \right),
\end{equation}
and the T--plural Dirichlet projector is\begin{equation} \wh Q=\left(
                \begin{array}{ccc}
                  \frac{1}{2} & -\frac{1}{2}  & 0 \\
                  -\frac{1}{2} & \frac{1}{2} & 0 \\
                  0& 0 & 1 \\
                \end{array}
              \right) .
\end{equation} The conditions (\ref{allconds}) are satisfied and the
condition (\ref{dm151}) implies both eq.\ (\ref{origbc}) and eq.\ (\ref{pluralbc}). The matrix (\ref{hathphidm1}) has
eigenvalues $(-1,-1,1)$ where $+1$ corresponds to a spacelike direction. Hence we get a Euclidean D0--brane. Similarly
to the D2--brane case also here neither $\gluingm$ nor $\gluingmh$ depends on elements of the groups $G$ and $\hat G$.
\end{itemize}

\subsection{Gluing matrices that produce $\wh \gluingmgen$ dependent only on $\wh
G$}\label{constC} The lesson we have learned from the previous subsection is that in some cases the \tfn{} of
coordinates (\ref{hatphi1}) may cure the problem of dependence of the gluing matrix $\wh \gluingmgen$ on elements of
the group $\bar G$. In particular, in our three-dimensional example it turned out that if D0-branes in $Bianchi\,5$
are contained in the hypersurface of constant $y^2$ located at $y^2=0$ then due to eqs.\ (\ref{hatphi1}) the plural
gluing matrices are well--defined.

In the present section we address the problem of coordinate cross--dependence from another point of view. We shall
assume that the plural gluing matrix depends on elements of $\hat G$ only, i.e., it is independent of the dual
coordinates on $\bar G$, and we derive the gluing matrices on both sides of the plurality that make this assumption
possible. Inspecting the transformation formula (\ref{gluingmh}) for the gluing operator we find that the T--plural
gluing matrix $\gluingmh$ is a function on $\wh G$ if and only if the matrix--valued function $$C(g) =F^{-t}(g) \cdot
\gluingm(g) \cdot F(g)$$ extended to a function on the whole Drinfel'd double as $C_D(l)=C(g)$ where $l=g\tilde h$
satisfies
\begin{equation} \label {aaa} C_D({\hat g} \bar h)=C_D(\hat g).\end{equation}

In our particular setting, where the relations between original and new coordinates on the \dd{} $D$ are given by
eqs.\ (\ref{hatphi1}) and (\ref{hprime}), we find that (only) the following combinations of $\widehat{y}$'s can be
written in terms of the original $y$'s,
$$
\widehat{y}^2-\widehat{y}^1=2 y^3, \qquad \widehat{y}^3=-y^1. $$
Consequently, if the original gluing matrix has the form
\begin{equation}\label{constH}
    \gluingm(g)=F^t(g)\cdot C \cdot F\-1(g),
\end{equation}
where $C=C(y^1,y^3)$, then the gluing matrices $\gluingmh$ given by eq.\ (\ref{gluingmh}) and $\wh {\gluingmphi},\,\wh
{\gluingmlambda}$ given by eq.\ (\ref{hrphi}), can be expressed as functions on $\wh G$ only, i.e., they are
well--defined. The condition (\ref{RgRg}) that $\gluingm$ of the form (\ref{constH}) preserve the metric yields
\begin{equation}\label{cece}
    C\cdot(E_0^{-1}+E_0^{-t})\cdot C^t=(E_0^{-1}+E_0^{-t}).
\end{equation}
In other words, the matrices $C$  belong to the representation of the group $O(n, \dim G - n)$ given by the constant
symmetric matrix $(E_0^{-1}+E_0^{-t})$ with signature $n$.

For $E_0$ of the form (\ref{E0}) we get the following possibilities,
\begin{eqnarray}
  C &=& \left(\matrix{- \frac{{\alpha }^2}{2\, \beta } & \alpha  & \beta  \cr - \epsilon
  \,
\frac{\alpha }{\beta }  & \epsilon& 0 \cr \frac{1}{\beta } & 0 & 0 \cr  }  \right) , \label{C1} \\
  C &=& \left(
\matrix{ \frac{{\left( \alpha+\epsilon  \right) }^2}{4\, \beta } & \frac{ \ 1 - {\alpha }^2  }{2\, \gamma } & -\frac{
{\left( \alpha-\epsilon \right) }^2\, \beta }{2\, {\gamma }^2} \cr -\frac{ \left(  \alpha+\epsilon \right) \, \ \gamma
}{2\, \beta } & \alpha & \frac{\left( \alpha - \epsilon \right) \, \beta }{\gamma } \cr \
-\frac{{\gamma }^2}{2\, \beta } & \gamma  & \beta  \cr  }  \right), \label{C2} \\
  C &=& \left(
\matrix{ \frac{1}{\beta } & \alpha  & -\frac{ {\alpha }^2\, \beta }{2} \cr 0 & \epsilon & - \epsilon\,\alpha \, \beta
\cr 0 & 0 & \beta \ \cr } \right) , \label{C3}
\end{eqnarray}
where $\epsilon =\pm 1,$ and  $\alpha, \beta, \gamma$ are arbitrary functions of $y^1$ and $y^3$. In addition, the
matrices $\gluingmphi$ and $\wh{\gluingmphi}$ calculated from eqs.\ (\ref{hrphi}), (\ref{gluingmh}), (\ref{constH})
must satisfy the conditions (\ref{allconds}) so that further restrictions on the matrices $C$ are imposed.
\begin{itemize}
    \item Case (\ref{C1}): The conditions (\ref{allconds}) for
    $\gluingmgen$
    are satisfied  only if $\alpha=0$. The gluing matrices then read
\begin{equation}\label{HRC1}
   \gluingm= \left(
\begin{array}{ccc}
 0 & 0 & \frac{1}{\beta}  \\
 0 & \epsilon & 0 \\
 \beta& 0 & 0
\end{array}
\right),\ \ \epsilon =\pm 1 ,\ \ \beta=\beta(y^1,y^3) ,\end{equation}
\begin{equation}\gluingmh=\half\ \left(
\begin{array}{ccc}
 -{\epsilon}& -{\epsilon}& {\beta}\\
 -{\epsilon}& -{\epsilon}& -{\beta}\\
 \frac{2}{\beta}  & -\frac{2}{\beta}  & 0
\end{array}
\right),\ \ \epsilon =\pm 1 ,\ \ \beta=\beta(-\wh y^3,\frac{\wh y^2-\wh y^1}{2}).
\end{equation}The conditions (\ref{allconds}) are satisfied for $\wh{\gluingmgen}$ as well. This corresponds to the \tfn{} of D1--branes to
D0--branes for $\epsilon=+1$, and D0--branes to D1--branes for $\epsilon=-1$.
\item { Case (\ref{C2}): The conditions
(\ref{allconds}) are satisfied for $\gluingmgen$ only if $\alpha=-\epsilon-2\beta$. The gluing matrices then
read
\begin{equation}\gluingm=\left(
\begin{array}{ccc}
 \beta & \gamma & -\frac{\gamma^2}{2 \beta} \\
 -\frac{2 (\beta+\epsilon) \beta}{\gamma} & -\epsilon-2 \beta &
   \gamma \\
 -\frac{2 (\beta+\epsilon)^2 \beta}{\gamma^2} & -\frac{2
   (\beta+\epsilon) \beta}{\gamma} & \beta
\end{array}
\right),\; \beta=\beta(y^1,y^3),\gamma=\gamma(y^1,y^3)\end{equation} \begin{equation}{\gluingmh}=\left(
\begin{array}{ccc}
 \frac{(\beta+\epsilon) \beta \kappa ^2+\gamma (3 \beta+\epsilon)
   \kappa +2 \gamma^2}{2 \kappa  \gamma} &
   \frac{(\beta+\epsilon) \beta \kappa ^2+\gamma (\beta+\epsilon)
   \kappa -2 \gamma^2}{2 \kappa  \gamma} & -\frac{(2
   \gamma+\kappa  (\beta+\epsilon)) (\beta+\epsilon) \beta}{\kappa
    \gamma^2} \\
 \frac{-(\beta+\epsilon) \beta \kappa ^2+\gamma (\beta+\epsilon)
   \kappa +2 \gamma^2}{2 \kappa  \gamma} &
   \frac{-(\beta+\epsilon) \beta \kappa ^2+\gamma (3
   \beta+\epsilon) \kappa -2 \gamma^2}{2 \kappa  \gamma} &
   \frac{(\kappa  (\beta+\epsilon)-2 \gamma) (\beta+\epsilon)
   \beta}{\kappa  \gamma^2} \\
 -\frac{\gamma (\gamma+\kappa  \beta)}{2 \beta} &
   \frac{\gamma (\gamma-\kappa  \beta)}{2 \beta} &
   \beta
\end{array}
\right), \end{equation}
where $\beta=\beta(-\wh y^3,\frac{\wh y^2-\wh y^1}{2}),\gamma=\gamma(-\wh y^3,\frac{\wh y^2-\wh y^1}{2}) $.
For $\epsilon=-1$  the dependence of $\beta$ and $\gamma$ on $y^1, y^3$ is
constrained by the condition (\ref{Liepiinteg1}) that yields \begin{equation}\label{conintc2}
e^{y^1}\gamma^2\left(\gamma\frac{\partial \beta}{\partial
   y^3}-\beta\frac{\partial \gamma}{\partial
   y^3}\right) = 2\beta^2\left(\gamma\frac{\partial \beta}{\partial
   y^1}+\frac{\partial \gamma}{\partial
   y^1}-\beta\frac{\partial \gamma}{\partial
   y^1}\right).\end{equation}  For $\epsilon=1$ we do not get any constraint on the functions $\beta,
   \gamma$.

The condition (\ref{LiepiEpiR}) is not satisfied for the matrix $\wh{\gluingmgen}$ unless we replace
$\widehat\cf $ by $\widehat\cf +\widehat\Delta$, where $\widehat N\widehat\Delta\widehat
N^t=\widehat\Delta$.

For $\epsilon=1$ \begin{equation}\widehat\Delta =\left(
\begin{array}{ccc}
 0 & -\frac{\beta}{\gamma} & -\frac{\gamma\, e^{-{\wh y}^3}}{2 + 2\,\beta } \\
 \frac{\beta}{\gamma} & 0 &  -\frac{\gamma\, e^{-{\wh y}^3}}{2 + 2\,\beta } \\
  \frac{\gamma\, e^{-{\wh y}^3}}{2 + 2\,\beta }  &  \frac{\gamma\, e^{-{\wh y}^3}}{2 + 2\,\beta } & 0
\end{array}
\right)\end{equation} and it is closed along the branes for arbitrary $\beta,\gamma$. This case
corresponds to the \tfn{} of D0--branes to D1--branes.

For $\epsilon=-1$ \begin{equation}\widehat\Delta =\left(
\begin{array}{ccc}
 0 & -\frac{\beta-1}{\gamma} & -\frac{\gamma\, e^{-{\wh y}^3}}{2\,\beta } \\
 \frac{\beta-1}{\gamma} & 0 &  -\frac{\gamma\, e^{-{\wh y}^3}}{2\,\beta } \\
  \frac{\gamma\, e^{-{\wh y}^3}}{ 2\,\beta }  &  \frac{\gamma\, e^{-{\wh y}^3}}{ 2\,\beta } & 0
\end{array}
\right)\end{equation} and it is closed along the branes due to (\ref{conintc2}). This case  corresponds to
the \tfn{} of D1-branes to D2-branes.
\item Case (\ref{C3}): The conditions (\ref{allconds})
for both $\gluingmgen$ and $\wh{\gluingmgen}$ are satisfied if $\beta =\pm 1$ and $ \alpha=0$. This corresponds to the
\tfn{} of D2--branes to D1--branes and D(-1)--branes to D0--branes,
\begin{equation} \gluingm=\pm\left(
                \begin{array}{ccc}
                  1 & 0 & 0 \\
                  0 & 1 & 0 \\
                  0& 0 & 1 \\
                \end{array}
              \right), \ \ \ \gluingmh=\pm\left(
                \begin{array}{ccc}
                  0 & -1 & 0 \\
                  -1 & 0& 0 \\
                  0& 0 & 1 \\
                \end{array}
              \right),
\end{equation} as presented in Section \ref{Dbranes}, and
\begin{equation} \gluingm=\pm\left(\matrix{ 1 & 0 & 0 \cr 0 & -1 & 0 \cr 0 & 0 & 1 \cr  }\right), \ \ \
\gluingmh=\pm\left(\matrix{ 1 & 0 & 0 \cr 0 & 1 & 0 \cr 0 & 0 & 1 \cr }\right),\end{equation} which correspond to \tfn
s of D1--branes to D2--branes and of D0--branes to D(-1)--branes.

Besides that the conditions (\ref{allconds}) for $\gluingmgen$  are also satisfied for
\begin{equation}\beta=-\epsilon, \qquad \frac{\partial\alpha}{\partial y^1}=0 \label{intconc3}.\end{equation}
However, to satisfy the condition (\ref{LieRQ}) for $\wh\gluingmgen$, i.e. $\wh\gluingmgen \cdot \wh Q = \wh Q \cdot \wh \gluingmgen = -\wh Q$
 with $\beta=-\epsilon= -1$ we must set
$\alpha=0$. Thus for
$\beta=-\epsilon=-1$ we see that in general (i.e., for $\alpha \neq 0$) the \pltp y does not
preserve the condition (\ref{LieRQ}).

 If $\alpha \neq 0$ and $\beta=-\epsilon= 1$ then we have $\wh Q=0$ and the condition (\ref{LieRQ}) holds trivially. We can satisfy the condition 
(\ref{LiepiEpiR}) by replacing $\widehat\cf $ by $\widehat\cf
+\widehat\Delta$ where \begin{equation}\widehat\Delta =\left(
\begin{array}{ccc}
 0 & \half \alpha & 0  \\
 -\half \alpha  & 0 &0  \\
 0 &  0 & 0
\end{array}
\right).\end{equation} This form is closed 
due to (\ref{intconc3}). The gluing matrices in this case \begin{equation}
\gluingm=\left(
\begin{array}{ccc}
 1 & 0 & 0 \\
 \alpha  & -1 & 0 \\
 -\frac{\alpha ^2}{2} & \alpha  & 1
\end{array}
\right),\ \ \ \gluingmh= \left(
\begin{array}{ccc}
 1-\frac{\alpha  \kappa }{4} & -\frac{\alpha  \kappa }{4} & \frac{\alpha }{\kappa }-\frac{\alpha
   ^2}{4} \\
 \frac{\alpha  \kappa }{4} & \frac{\alpha  \kappa }{4}+1 & \frac{\alpha  (\alpha  \kappa +4)}{4
   \kappa } \\
 0 & 0 & 1
\end{array}
\right).\end{equation} correspond to the \tfn{} of D1--branes to
D2--branes.}\end{itemize}

We remark that in three dimensions the integrability condition (\ref{Liepiinteg1}) is nontrivial only if the rank of
the Neumann projector $N$ is equal to two; otherwise the distribution  $\Delta={\rm Im}\, (N)$ is integrable on
dimensional grounds. In two dimensions, investigated below, the condition (\ref{Liepiinteg1}) is always trivially
satisfied.

\section{Two--dimensional example} \label{2dimex}
The only \sm s with two--dimensional targets that can be transformed under T--plurality with nonisomorphic
decompositions of  a \dd{} are generated by the semi--Abelian four--dimensional \dd{} of \cite{hlasno:pltdm2dt}. It
has decompositions into two different pairs of maximally isotropic Lie subalgebras, namely the semi--Abelian Manin
triple with basis $T_1,T_2,\tilde T^1,\tilde T^2$ and Lie brackets (only nontrivial brackets are displayed)
\begin{equation} \label{sa} [T_1,T_2]=T_2,\ [\tilde T^2,T_1]=\tilde T^2,\ [\tilde T^2,T_2]=-\tilde T^1 ,
\end{equation} and the so--called type B nonabelian Manin triple with basis $\hat T_1,\hat T_2,\bar T^1,\bar
T^2$ and Lie brackets
\[ [\hat T_1,\hat T_2]=\hat T_2,\ \ [\bar T^1,\bar T^2]=\bar T^1, \]
\begin{equation} \label{tB} [\hat T_1,\bar T^1]= \hat T_2,\ \ [\hat T_1,\bar T^2]=-\hat T_1-\bar T^2,\ \ [\hat T_2,\bar T^2]=\bar
T^1. \end{equation} A simple transformation between the bases of these two decompositions is given by
\begin{eqnarray}
\nonumber \hat T_1   =  -{T}_1+{T}_2,   &&
 \hat T_2  =  \tilde T^1+\tilde T^2, \\
\bar T ^1  =  \tilde T^2, && \bar{T}^2  =  {T}_1 ,
\end{eqnarray}
which corresponds to the transformation matrix \begin{equation} \left(\matrix{\PP&\QQ \cr \RR&\SSS \cr} \right)
=\left(
\begin{array}{cccc}
 0 & 0 & 0 & 1 \\
 1 & 0 & 0 & 1 \\
 0 & 1 & -1 & 0 \\
 0 & 0 & 1 & 0
\end{array}
\right). \label{CSASS}\end{equation} The coordinate transformation on the \dd{} that follows from this reads
\begin{eqnarray}
\hat x^1 = -\ln(-x^2+1),& &
  \hat x^2 = -\frac{\tilde x^1}{x^2-1},\nonumber
  \\
\bar x^1 = \frac{\tilde x^1 \exp(x^1)+x^2 \tilde x^2-\tilde x^2}{x^2-1},&&\bar x^2 = -\ln(-x^2+1)+x^1.
\end{eqnarray}
We shall consider examples of two--dimensional \sm s given by the matrices \begin{equation} E_0=\left(\matrix{1&0 \cr
0&\kappa \cr} \right),\ \ \wh E_0=\left(\matrix{\kappa&1 \cr -1&1 \cr} \right),
\end{equation} where $\kappa$ is a real constant. The corresponding tensors $\cf$, $\hat \cf$ are calculated
from eqs.\ (\ref{Fg}) and (\ref{Fghat}), where
\begin{equation} g=\exp(x^2T_2)\,\exp(x^1T_1),\ \ \hat g=\exp(\hat x^2\hat T_2)\,\exp({\hat x}^1\hat T_1).
\end{equation} They read \begin{equation} \cf(x^\mu)=\left(
\begin{array}{cc}
 \kappa\,  (x^2)^2+1 & -\kappa \, x^2 \\
 -\kappa \, x^2 & \kappa
\end{array}
\right),\label{fsa}\end{equation}
\begin{equation} \wh\cf(\hat x^\mu)=\frac{1}{{-2 e^{{\hat x^1}} \kappa +\kappa +e^{2 {\hat x^1}} (\kappa
+1)}}\left(
\begin{array}{cc}
  {(\hat x^2)^2+\kappa } & - \kappa
   +e^{{\hat x^1}} (\kappa +1)-{\hat x^2}  \\
  {\kappa -e^{{\hat x^1}} (\kappa +1)-{\hat x^2}} &  {1}
\end{array}
\right). \label{fss}\end{equation}
 Unfortunately, the metrics of both models are curved and we are not able to solve the equations
of motion. Nevertheless, we can at least find the gluing matrices that satisfy the conditions (\ref{allconds}).
Moreover, we require that the gluing matrices depend only on the coordinates where the \sm s live so that we have to
take $\gluingm$ of the form (\ref{constH}), with $C$ depending only on $x^2$ in order
to satisfy eq.\ (\ref{aaa}).

The \cond{} (\ref{cece}) restricts $C$ to the form
\begin{equation}
  C = \left(
\begin{array}{cc}
 \epsilon_1\sqrt{1-\gamma ^2 \kappa } & \epsilon_2\gamma  \kappa  \\
 \gamma  & -\epsilon_1\epsilon_2\sqrt{1-\gamma ^2 \kappa }
\end{array}
\right),\label{Csa1}\end{equation}
 where $\gamma$ is an arbitrary function of $x^2$ and $\epsilon_1,\epsilon_2=\pm1$.
The conditions (\ref{LieRQ}), (\ref{LiepigQ}) are then satisfied for
all corresponding matrices $\gluingmgen$. The condition
(\ref{LiepiEpiR}) is satisfied only if $\epsilon_2=1$ or
$\epsilon_2=-1, \gamma=0$.

If $\epsilon_2=-1, \gamma=0$ then the conditions (\ref{allconds})
are satisfied for the transformed \sm{} as well. The gluing matrices
are
\begin{equation} \gluingm= \left(
\begin{array}{cc}
 \epsilon_1& 0  \\
 0  & \epsilon_1
\end{array}
\right),\ \ \gluingmh= \left(
\begin{array}{cc}
 \epsilon_1& 0  \\
 0  & -\epsilon_1
\end{array}
\right),\end{equation} so that the boundary conditions for the \sm{}
on $G$ are purely Dirichlet or purely Neumann. Interpretation of the
boundary condition for the \sm{} on $\hat G$ as either usual
D0--branes or Euclidean (spacelike) D1--branes depends on the
signature of the metric, i.e., on the sign of $\kappa$.

{ If $\epsilon_2=1$ then \begin{equation}
 \gluingm=  \left(
\begin{array}{cc}
  -\epsilon_1\epsilon_2\sqrt{1-\gamma ^2 \kappa } & \gamma   \\
 \epsilon_2\gamma  \kappa  & \epsilon_1\sqrt{1-\gamma ^2 \kappa }
\end{array}
\right),\end{equation}  The transformed gluing matrix $\gluingmh$ is easily obtained from (\ref{gluingmh}) but it is too complicated
to display here.
The conditions (\ref{LieRQ}),
(\ref{LiepigQ}) are satisfied for all these matrices $\gluingmh$.
The condition (\ref{LiepiEpiR}) can be always satisfied by replacing
$\widehat\cf $ by $\widehat\cf +\widehat\Delta$ where \begin{equation}
\widehat\Delta =\left(
\begin{array}{cc}
  0 & \widehat\Delta_{12} \\
 -\widehat\Delta_{12}&0
\end{array}
\right),\end{equation}
\begin{equation}\widehat\Delta_{12}=\frac{1 + \gamma \,\kappa  -
\epsilon_1\,
     {\sqrt{1 - {\gamma }^2\,\kappa }} +
    e^{\hat x^1}\,
     \left( \gamma \,\left( 1 - \kappa  \right)  +
       2\,\epsilon_1\,{\sqrt{1 - {\gamma }^2\,\kappa }} \right) }{
     \gamma \,\kappa  + e^{2\,\hat x^1}\,
     \left( \gamma \,\left( -1 + \kappa  \right)  -
       2\,\epsilon_1\,{\sqrt{1 - {\gamma }^2\,\kappa }} \right)  +
    2\,e^{\hat x^1}\,
     \left( -\left( \gamma \,\kappa  \right)  +
       \epsilon_1\,{\sqrt{1 - {\gamma }^2\,\kappa }} \right) } \end{equation}

In the case when the denominator of $\widehat\Delta_{12}$ vanishes i.e.
for $\hat x^1$ satisfying (recall that $\gamma$ is a function of
$\hat x^1$)\begin{equation}\label{hatx1} \frac{\epsilon_3 -\epsilon_1 \gamma
\,\kappa + {\sqrt{1 - {\gamma }^2\,\kappa }}}
  {\epsilon_1\gamma \,\left( 1 - \kappa  \right)  +
    2\,{\sqrt{1 - {\gamma }^2\,\kappa }}}= e^{\hat x^1}\end{equation} we get \begin{equation}
\gluingmh=-\epsilon_1\epsilon_3 \left(
\begin{array}{cc}
 1& 0  \\
 0  & 1
\end{array}
\right). \end{equation}

The eigenvalues of $\gluingm$ are $+1,-1$ corresponding to either
usual D0--branes or Euclidean D1--branes, while the boundary
conditions for the \sm{} on $\hat G$ are purely Neumann except for
$\hat x^1$ satisfying (\ref{hatx1}) with $\epsilon_3=\epsilon_1$
in which case they are purely Dirichlet and $\widehat\Delta_{12}$ becomes singular\footnote{whereas when (\ref{hatx1}) holds with $\epsilon_3=-\epsilon_1$ we have $\gluingmh=\unit$ and the singularity of $\widehat\Delta_{12}$ is only apparent -- it becomes an expression of the form $\frac{0}{0}$ with a finite and well--defined limit}. }

\section{Conclusions} \label{Conclusions} We have derived a formula (\ref{gluingmh}) for
\tfn {} of boundary conditions under \pltp y. The examples in Section \ref{Dbranes} confirm that the formula works for
solutions of the \eqn s of motion of the \sm s. This is not surprising since it was derived using these \eqn s. The
problem is that the transformed gluing matrix may depend on elements of the original group (and hence on elements of
the dual group $\bar G$), so that only special forms of gluing matrices are transformable under \pltp y.

To ensure that the gluing matrices transformed by \pltp y depend only on the coordinates of the groups where the \sm s
live we can restrict them to the form (\ref{constH})
\[
    \gluingm=F^t(g)\cdot C \cdot F\-1(g).
\]
The matrix $C$ must be constant or depend only on a particular subset of coordinates on $G$ that transform into
coordinates on $\hat G$.

Another problem is that not all conditions (\ref{allconds}) for consistent D--branes are preserved under
\pltp y. We have proven that the condition (\ref{confcond}), i.e.,  $\gluingmgen \cdot {\cal G} \cdot
\gluingmgen^t = {\cal G}$, is always preserved. In Euclidean signature this implies the preservation of
conditions (\ref{LieRQ}) and (\ref{LiepigQ}), i.e., $\gluingmgen\cdot Q = Q\cdot \gluingmgen$ and $N \cdot
{\cal G} \cdot Q^t =0$. { As we have seen in the investigation of the matrix $C$ of the form (\ref{C3}) in
Section \ref{constC} it is not necessarily so in the case of indefinite signature. In that case the
transformed gluing matrix may become non--diagonalizable (in the sense of non--diagonal Jordan canonical
form) and consequently the projector on $(-1)$--eigenspace cannot satisfy $\gluingmgen\cdot Q = Q\cdot
\gluingmgen$.  Nevertheless, when such an obstruction did not arise the conditions (\ref{LieRQ}) and
(\ref{LiepigQ}) were satisfied in all cases investigated here also in the indefinite signature. Similarly,
the integrability condition (\ref{Liepiinteg1}) was preserved in all examples.

On the other hand, we have seen explicitly that the condition (\ref{LiepiEpiR}), i.e., $ N \cdot ({\cal F} - {\cal F}^t \cdot \gluingmgen^t ) =0 $ is not preserved in general under \pltp y and that in the transformed background it must be modified by the presence of an electric field constrained to the branes and interacting with oppositely charged endpoints of the string. We have moreover seen in several cases with nonconstant matrix $C$ in (\ref{constH}) that the closedness (\ref{cor_closure}) of this additional electric field  is intimately related to the integrability of the Neumann distribution (\ref{Liepiinteg1}) in the original model. It is an open question whether and how this behavior can be proven in general or whether it happens just in the low dimensions investigated here.}

\section{Acknowledgements} This work was supported by the project of the Grant Agency of the Czech Republic
No.\ 202/06/1480 and by the research plans LC527 15397/2005--31 and MSM6840770039 of the Ministry of Education of the
Czech Republic (L.H. and L.\v S.). C.A. acknowledges support by the Japanese Society for the Promotion of Science
(JSPS).

\end{document}